\documentclass[runningheads,a4paper]{llncs}

\usepackage{amssymb}
\usepackage{graphicx}

\usepackage{url}
\urldef{\mailsa}\path|{alfred.hofmann, ursula.barth, ingrid.haas, frank.holzwarth,|
\urldef{\mailsb}\path|anna.kramer, leonie.kunz, christine.reiss, nicole.sator,|
\urldef{\mailsc}\path|erika.siebert-cole, peter.strasser, lncs}@springer.com|
\newcommand{\keywords}[1]{\par\addvspace\baselineskip
\noindent\keywordname\enspace\ignorespaces#1}

\usepackage{amssymb}
\usepackage{amsmath,bm}
\usepackage{algorithm}
\usepackage{algorithmic}
\usepackage{graphicx}

\newcounter{DefNum}
\usecounter{DefNum}
\newcounter{ExaNum}
\usecounter{ExaNum}
\newcommand{\Def}{\noindent\textbf{Definition~\arabic{DefNum}.~}\refstepcounter{DefNum}}

\graphicspath{{./fig/}}
\newcommand{\sims}{s}
\renewcommand\figurename{Figure}
\begin{document}

\mainmatter  % start of an individual contribution

\title{Semi-supervised evidential label propagation algorithm for graph data}

\titlerunning{Semi-supervised evidential label propagation}


%\author{Alfred Hofmann%
%\thanks{Please note that the LNCS Editorial assumes that all authors have used
%the western naming convention, with given names preceding surnames. This determines
%the structure of the names in the running heads and the author index.}%
%\and Ursula Barth\and Ingrid Haas\and Frank Holzwarth\and\\
%Anna Kramer\and Leonie Kunz\and Christine Rei\ss\and\\
%Nicole Sator\and Erika Siebert-Cole\and Peter Stra\ss er}


%\institute{Springer-Verlag, Computer Science Editorial,\\
%Tiergartenstr. 17, 69121 Heidelberg, Germany\\
%\url{http://www.springer.com/lncs}}


\author{Kuang Zhou\inst{1,2} \and Arnaud Martin\inst{2}
 \and Quan Pan\inst{1}}
\authorrunning{Kuang Zhou et al.} % abbreviated author list (for running head)
%Craig Chambers, Kim B. Bruce, and Elisa Bertino}
%
\institute{Northwestern Polytechnical University, \\ Xi'an, Shaanxi 710072, PR China
\and
DRUID, IRISA, University of Rennes 1, Rue E. Branly, 22300 Lannion, France
\\
kzhoumath@163.com,  Arnaud.Martin@univ-rennes1.fr, quanpan@nwpu.edu.cn}

\toctitle{Lecture Notes in Computer Science}
\tocauthor{Authors' Instructions}
\maketitle

\begin{abstract}
In the task of community detection, there often exists some useful prior information.  In this paper,  a Semi-supervised
clustering approach using a new Evidential Label Propagation strategy (SELP) is proposed to incorporate the domain knowledge
into the community detection model. The main advantage of SELP is that it can take limited supervised
knowledge to guide the detection process. The prior information of community labels
is expressed in the form of mass functions initially. Then a new evidential label propagation rule
is adopted to propagate the labels from labeled data to unlabeled ones. The
outliers can be identified to be in a special class. The experimental results  demonstrate
the effectiveness of SELP.

\end{abstract}

\section{Introduction}
With the increasing size of networks in real world,  community detection approaches should be  fast and accurate.
The Label Propagation Algorithm (LPA) \cite{raghavan2007near} is known to be one of the  near-linear solutions and benefits of easy implementation,
thus it forms a good basis for efficient community detection
methods. The behavior of LPA is not stable  because of the randomness. Different communities may be detected in
different runs over the same network. Moreover,  by assuming
that a node always adopts the label of the majority of its
neighbors, LPA ignores any other structural information existing in the neighborhood.

Semi-supervised classification has been widely studied for  classical data sets, but
there has been little work on semi-supervised community detection. In many scenarios  a
substantial amount of prior knowledge about the graph structure may be available.
It can reflect the application-specific knowledge about
cluster membership to some extent. For instance, in a co-authorship community network,
it may be possible to label a small subset of scholars based on their research interests.  In a social network application,
it may be desirable to label some nodes according to their affinity to some products.

In \cite{liu2014semi} the authors considered the individual labels as prior knowledge, {\em i.e.}
the true community assignments of certain nodes are known in advance. In their work the traditional
LPA is adapted, allowing a few nodes to have true community labels, but
the rest nodes are unlabeled. In face the presented semi-supervised community detection approach is an application of
the semi-supervised classification algorithm proposed by \cite{wang2008label} on graph data sets.

In this paper, we enhance the original LPA by introducing  new update and propagation strategies using the theory of belief functions.
The Semi-supervised version of Evidential Label Propagation (SELP) algorithm is presented.
SELP can take advantage of the limited amount of supervised information and consequently improve the detection results.

The remainder of this paper is organized as follows. Some basic knowledge is  briefly introduced in Section \ref{background}. The SELP algorithm is presented in detail in Section \ref{SELPpresent}. In order to show the effectiveness of the proposed
community detection approach, in Section \ref{secexp}  SELP algorithm is tested
 on different graph data sets.
Conclusions are drawn in the final section.

\section{Background}
\label{background}
In this section some related preliminary  knowledge will be presented. Some basis of belief function theory will
be recalled first. As this work is inspired from the LPA \cite{raghavan2007near} and E$K$-NNclus \cite{denoeux2015ek} clustering,
the two algorithms will also be briefly introduced.
\subsection{Theory of belief functions}
Let $\Omega=\{\omega_{1},\omega_{2},\ldots,\omega_{c}\}$ be the finite domain of
$X$, called the discernment frame. The belief functions are defined on the power
set $2^{\Omega}=\{A:A\subseteq\Omega\}$.

%\begin{definition}
The function $m:2^{\Omega}\rightarrow[0,1]$ is said to be the Basic Belief
Assignment (bba) on $2^{\Omega}$, if it satisfies:
\begin{equation}
\sum_{A\subseteq\Omega}m(A)=1.
\end{equation}
Every $A\in2^{\Omega}$ such that $m(A)>0$ is called a focal element.
%\end{definition}
The credibility and plausibility functions are defined  in Eqs.$~\eqref{bel}$ and $\eqref{pl}$ respectively:
\begin{equation}
Bel\text{(}A\text{)}=\sum_{B\subseteq A, B \neq \emptyset} m\text{(}B\text{)} ~~\forall A\subseteq\Omega,
\label{bel}
\end{equation}

\begin{equation}
 Pl\text{(}A\text{)}=\sum_{B\cap A\neq\emptyset}m\text{(}B\text{)},~~\forall A\subseteq\Omega.
 \label{pl}
\end{equation}
Each quantity $Bel(A)$  measures the total support given to $A$, while $Pl(A)$ represents potential amount of support to $A$.

If bbas $m_j, j=1,2,\cdots,S$ describing $S$
distinct items of evidence on $\Omega$, the DS
rule of combination \cite{ds2} of $S$ bbas can be mathematically defined as
\begin{align}\label{dsrule}
  %m_{1 \oplus 2} (A)
   (m_1\oplus m_2 &\oplus\cdots \oplus m_S)(X)=  \nonumber \\ &\begin{cases}
    0 & \text{if}~ X = \emptyset,\\
    %\frac{m_{1 \text{\textcircled{\scalebox{0.8}{\tiny{$\cap$}}}} 2} (A)}{1-m_{1 \text{\textcircled{\scalebox{0.8}{\tiny{$\cap$}}}} 2} (\emptyset)} &
     \frac{\sum\limits_{Y_1 \cap \cdots \cap Y_S = X} \prod_{j=1}^{S}m_j(Y_j)}{1-\sum\limits_{Y_1 \cap \cdots \cap Y_S = X} \prod_{j=1}^{S}m_j(Y_j)} & \text{otherwise}.
  \end{cases}
\end{align}
\subsection{E$K$-NNclus clustering}
Recently, a new decision-directed clustering algorithm for relational data sets, named E$K$-NNclus, is put forward based on
the evidential $K$ nearest-neighbor (E$K$-NN) rule~\cite{denoeux2015ek}.
Starting from an initial partition, the algorithm, called E$K$-NNclus, iteratively reassigns objects to clusters using
the E$K$-NN rule \cite{denoeux1995k}, until a stable partition is obtained. After convergence, the cluster membership of each
object is described by a  mass function assigning a mass to each specific cluster and to the whole
set of clusters.
\subsection{Label propagation}
Let $G(V,E)$ be an undirected network, $V$ is the set of  $N$ nodes, $E$ is the set of edges. Each node $v(v \in V)$ has a label $c_v$.
Denote by $N_v$ the set of neighbors of node $v$.
The Label Propagation Algorithm (LPA) uses the network structure alone to guide its
process. It starts from  an initial configuration where every node has a unique label. Then at every step one node (in
asynchronous version) or each node (in a synchronous version) updates
its current label to the label shared by the maximum number of its neighbors. For node $v$, its new label
can be updated to $\omega_j$ with
\begin{equation}
  j = \arg \max_{l} \{|u:c_u=l,u \in N_v| \},
\end{equation}
where  $|X|$ is the cardinality of set $X$, and $N_v$ is the set of node $v$'s neighbors. When there are multiple
maximal labels among the neighbors’ labels, the new label is
picked randomly from them.  By this iterative process
densely connected groups of nodes form consensus on one label to form communities, and each node has more neighbors in its own
community than in any of other community.    Communities
are identified as a group of nodes sharing the same label.
\section{Semi-supervised label propagation}
\label{SELPpresent}
Inspired from LPA and E$K$-NNclus \cite{denoeux2015ek}, we propose here SELP algorithm for graph data sets with prior information.
The problem of semi-supervised community detection will be first described in a mathematical way, and then
the proposed SELP algorithm will be presented in detail.

%Inspired from LPA and E$K$-NNclus, we propose here the SELP algorithm for community detection.
\subsection{Problem restatement and notions}
Let $G(V,E)$ denote the graph, where $V$ is the set of $n$ nodes and $E \subseteq V \times V$
is the set of edges.  Generally, a network can be expressed by its adjacent matrix  $\bm{A}=(a_{ij})_{n\times n}$, where $a_{ij}=1$ indicates that there is a direct edge between nodes $i$ and $j$, and 0 otherwise.

Assume that there are $c$ communities in the graph. The  set of labels is denoted by $\Omega = \{\omega_1, \omega_2, \cdots, \omega_c\}$.  In addition,
in order to make sure that the solution is unique, we assume that there must be at least one labeled vertex in each community. The $n$ nodes in
set $V$ can be divided into two parts: $$V_L=\{(n_1, y_1), (n_2, y_2), \cdots, (n_l, y_l)\},~~y_j \in \Omega$$ for the labeled nodes, and
$$V_U=\{n_{l+1}, n_{l+2}, \cdots, n_n\}$$ for the unlabeled ones.  The main task of the semi-supervised community detection is to make
models propagating the labels from nodes in $V_L$ to those in $V_U$, and further determine the labels of those unlabeled vertices.
\subsection{The dissimilarities between nodes}
\label{nodeinf}
Like the smooth assumption in the semi-supervised graph-based learning methods \cite{zhu2005semigraph}, here we assume
that the more common neighbors the two nodes share, the larger probability that they belong to the
same community. Thus in this work, the index considering the number of shared common neighbors  is adopted to measure the
similarities between nodes.

\Def Let the set of neighbors of node $n_i$ be $N_i$, and the degree of node $n_i$ be $d_i$. The similarity between nodes $n_i$ and
$n_j$ ($n_i, n_j \in V$) is defined as
\begin{equation}
  \sims_{ij} = \begin{cases}
    \frac{|N_i \cap N_j|}{d_i + d_j}, &~\text{if}~ a_{ij} = 1\\
    0, &~\text{otherwise}.
  \end{cases}
\end{equation}
Then the dissimilarities associated with the similarity measure can be defined as
\begin{equation}
  d_{ij} = \frac{1-\sims_{ij}}{\sims_{ij}}, ~ \forall ~ n_i, n_j \in V.
\end{equation}

\subsection{Evidential label propagation}

For a labeled node $n_j \in V_L$ in community $\omega_k$, the initial bba can be  defined as a Bayesian categorical mass function:
\begin{equation}\label{massVL}
  m^{j}(A) = \begin{cases}
    1 & ~ \text{if}~A = \{\omega_k\} \\
    0 & ~ \text{otherwise}.
  \end{cases}
\end{equation}
For an unlabeled node $n_x \in V_U$, the vacuous  mass assignment can be used to express our ignorance about its community label:
\begin{equation}\label{massVU}
  m^x(A) = \begin{cases}
    1 & ~ \text{if}~A = \Omega \\
    0 & ~ \text{otherwise}.
  \end{cases}
\end{equation}

To determine the label of node $n_x$, its neighbors can be regarded as distinct information sources. If there are $|N_x| = r_x$ neighbors
for node $n_x$, the number of sources is $r_x$. The reliability of
each source depends on the similarities between nodes. Suppose that there is a neighbor $n_t$ with
label $\omega_j$, it can provide us with a bba describing the belief on  the community label of node $n_x$ as \cite{denoeux2015ek}
\begin{align}\label{massNei}
  m_t^x(\{\omega_t\}) &= \alpha * m^t(\{\omega_j\}) \nonumber, \\
  m_t^x(\Omega) &= m^t(\Omega) + (1-\alpha) * m^t(\{\omega_j\})\nonumber, \\
  m_t^x(A) &= 0, ~~\text{if}~A \neq \{\omega_j\}, \Omega,
\end{align}
where $\alpha$ is the discounting parameter such that $0 \leq \alpha \leq 1$.  It should be determined according to the similarity between
nodes $n_x$ and $n_t$. The more similar the two nodes are, the more reliable the source is. Thus $\alpha$ can be set as
a decreasing function of $d_{xt}$. In this work we suggest to use
\begin{equation}
  \alpha = \alpha_0 \exp\{-\gamma d_{xt}^\beta\},
\end{equation}
where parameters $\alpha_0$ and $\beta$ can be set to be 1 and 2 respectively as default, and $\gamma$ can be set to
\begin{equation}
  \gamma = 1/\text{median}\left(\left\{d_{ij}^\beta,~ i=1,2,\cdots,n, ~j \in N_i\right\}\right).
\end{equation}
After  the $r_x$ bbas from its neighbors are calculated using Eq.~\eqref{massNei}, the fused bba of node $n_x$ can be got by the use of
Dempster's combination rule:
\begin{equation}\label{combinemass}
  m^x = m_1^x \oplus m_2^x \oplus \cdots \oplus m_{r_x}^x.
\end{equation}

The label of node $n_x$ can be determined by the maximal value of $m^x$. The main
principle of semi-supervised learning is to take advantage of the
unlabeled data. It is an intuitive way to add  node $n_x$
(previously in set $V_U$ but  already be labeled now) to set $V_l$ to train the classifier. However, if the predicted label of $n_x$ is wrong,
it will have very bad effects on the accuracy of the following predictions. Here a parameter $\eta$ is introduced to
control the prediction confidence of the nodes that to be added in $V_l$.
If the maximum of $m^x$ is larger than $\eta$, it indicates that the belief about the community of node $n_x$ is high and the 
prediction is confident. Then we remove node $n_x$ in $V_U$ and add it to set $V_L$. On the contrary, 
if the maximum of $m^x$ is not larger than $\eta$, it means that we can not 
make a confident decision about the label of $n_x$ based on the current information. Thus the node $n_x$ 
should be remained in set $V_U$. This is the idea of self-training \cite{li2005setred}.

In order to propagate the labels from the labeled nodes to the unlabeled ones in the graph,  a classifier should be first trained using
the labeled data in $V_l$.  For each node $n_x$ in $V_U$, we find its direct neighbors and construct bbas through
Eq.~\eqref{massNei}. Then the fused bba
about the community label of node $n_x$ is calculated
by Eq.~\eqref{combinemass}. The subset of the unlabeled nodes, of which the maximal bba is larger
than the given threshold $\eta$,  are selected to augment the labeled data set. The predicted labels of these nodes are set to be the class assigned
with the maximal mass. Parameter $\eta$ can be set to 0.7 by default in practice.
%First we select a node in $V_l$, and without loss of generality suppose it is node $n_j$.  Then we check the neighbors
%of node $n_j$, and calculate its bba according to Eqs.~\eqref{massNei} and \eqref{combinemass}. If the maximal assignment for
%the neighbors is larger than $\eta$, this neighbor node is added to set $V_L$ and it can be partitioned into the community
%with maximal mass assignment.  After checking all the neighbors of $n_j$,
%we select another node in $V_l$ and repeat the process until all the nodes in $V_L$ are checked.

After the above update process, there may still be some nodes in $V_U$. For these nodes, we can find their neighbors that are
in $V_L$, and then use Eqs.~\eqref{massNei} and \eqref{combinemass} to determine their bbas.
%The whole algorithm of SELP can be summarised in Algorithm \ref{alg:methodSELP}.

%\begin{algorithm}\caption{\textbf{:}~~~ SELP  algorithm}\label{alg:methodSELP}
%\begin{algorithmic}
%\STATE {\textbf{Input:} Graph $G(V,E)$. The set of labeled node $V_L$, and
%the set of unlabeled node $V_U$}.
%\STATE{\textbf{Parameters:}
%	~\\$\eta$: the parameter to control the prediction confidence \\
%$\alpha_0, \beta$: the parameter to determine the discounting factor \\
%$MaxIts$: the maximal update steps \\
%$PercFul$: the percentage of the labeled data
%}
%\STATE {\textbf{Initialization:}\\
% (1). Initialize the bba of each node in the network using Eqs.~\eqref{massVL} and \eqref{massVU}. \\
% (2). Let $it = 0$
%			}
%\REPEAT
%\STATE{
% (1). For each  node $n_x \in V_U$, find its $r_x$ direct neighbors and construct $r_x$ bbas of $n_x$ using Eq.~\eqref{massNei}. \\
% (2). Calculate the fused bba of node $n_x$ by Eq.~\eqref{combinemass}. \\
% (3). If the maximum of mass assignment of $n_x$ is larger than $\eta$, move node $n_X$ from set $V_U$ to set $V_L$. \\
% (4). $it = it + 1$.
%}
%\UNTIL{The percentage of nodes in $V_L$ is larger than $PercFul$ or the maximal update step is reached.}
%
%\STATE{
%If there are still some labels in $V_U$,   update their bbas based on the information from the neighbors using Eqs.~\eqref{massNei} and \eqref{combinemass}.
%}
%\STATE{
%\textbf{Output:} The bba matrix $\bm{M} = \{m_i\},~ i = 1,2,\cdots, N$.
%}
%\end{algorithmic}
%\end{algorithm}
%\subsection{Application on relational data}

\section{Experiment}
\label{secexp}
In order to verify the efficiency and effectiveness of the proposed SELP algorithm, some experiments on graph data sets will be conducted in this section, and the results by the use of different methods will be reported. The semi-supervised
community detection algorithm using label propagation (SLP) \cite{liu2014semi} and the unsupervised label propagation algorithm
will be used to compare the performance. The parameters in SELP are all set to the default values in the experiments.

\subsection{Real world networks}

\textbf{A. Karate Club network.}
In this experiment we  tested on the widely used benchmark  in detecting community structures, ``Karate Club".
The network consists of 34  nodes and 78 edges representing the friendship among the members of the club. During the development, a dispute
arose between the club's administrator and instructor, which eventually resulted in the club  split into two smaller clubs. The first one was an
instructor-centered group covering 16 vertices: 1–-8, 11-–14, 17-–18, 20 and 22, while the second administrator centered group consisted of the
remaining 18 vertices.
\begin{center} \begin{figure}[!htp] \centering
		\includegraphics[width=0.45\linewidth]{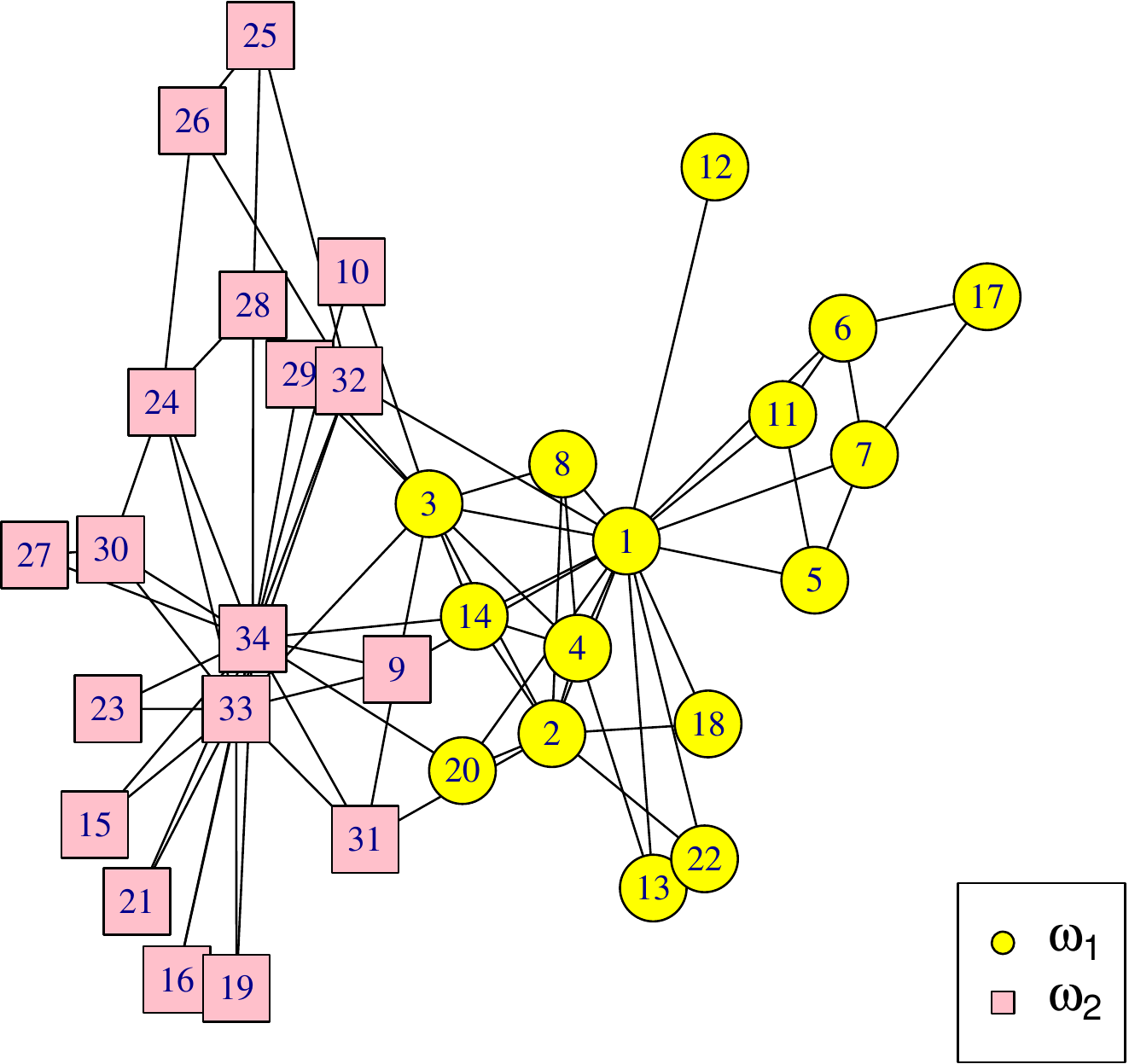}\hfill
\caption{Karate Club network.} \label{karateSELP} \end{figure} \end{center}

\begin{center} \begin{figure}[!htp] \centering
		\includegraphics[width=0.45\linewidth]{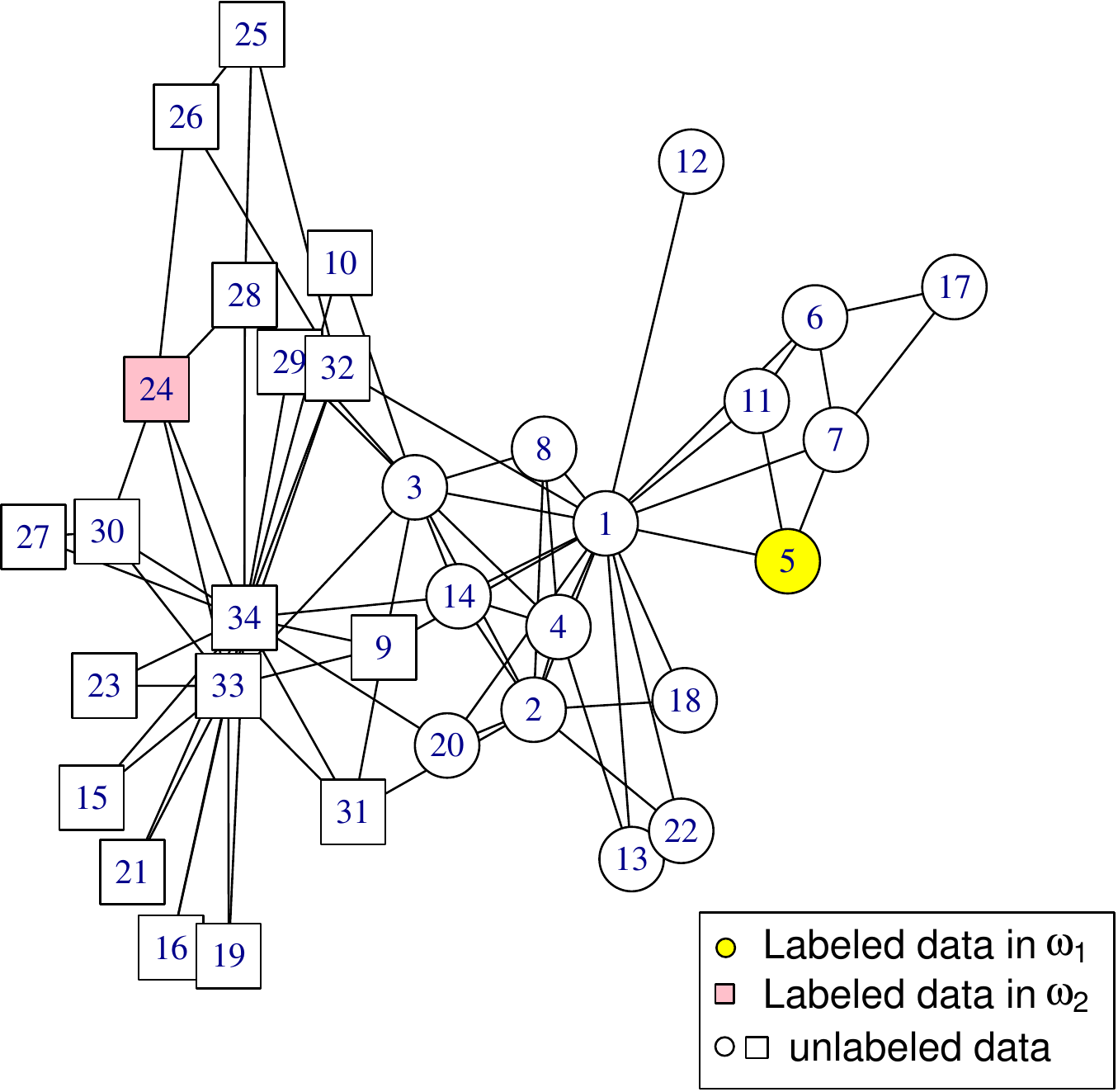}\hfill
        \includegraphics[width=0.45\linewidth]{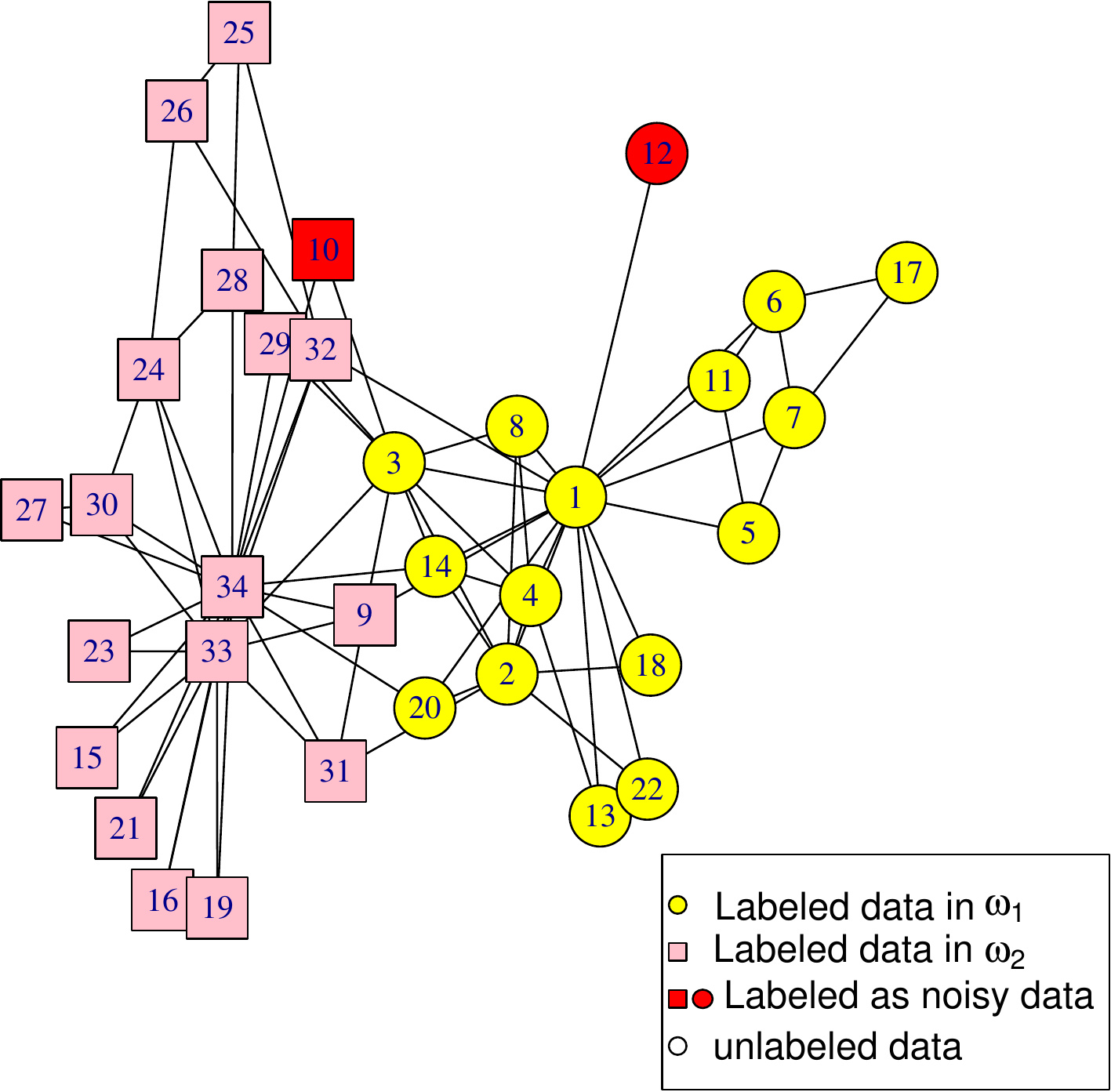} \hfill
        \parbox{.45\linewidth}{\centering\small a. Initialization} \hfill
		\parbox{.45\linewidth}{\centering\small b. $it = 5$}
\caption[The label propagation process on Karate Club network]{The results on Karate Club network. The nodes marked with color red are the
outliers detected by SELP.} \label{karate_label} \end{figure} \end{center}

\vspace{-2.5em}
In the first test, the labeled node in community $\omega_1$ was set to node 5, while that  in community $\omega_2$ was set to node 24.
After five steps, SELP algorithm stopped. The detailed update process is displayed in Figure \ref{karate_label}.
It can be seen from the figure that two outliers, nodes 10 and 12 are detected by SELP. From the original graph, we can
see that node 10 has two neighbors, nodes 3 and 34. But neither of them shares a common neighbor with node 10.
For node 12, it only connects to node 1, but has no connection with any other node
in the graph. Therefore, it is very intuitive that both the two nodes are regarded as outliers of the graph.

The detection results on  Karate Club network by SELP and SLP algorithms with different labeled nodes are shown in Table \ref{karate_table}.
The labeled vertices and its corresponding misclassified vertices are clearly presented in the table.
As can be seen from the table, nodes 10 and 12 are
detected as outliers in all the cases by SELP, and the two communities can be correctly classified most of the time. The performance
of SLP is worse than that of SELP when there is only one labeled data in each community.
For the nodes which are connected to both communities and located in the
overlap, such as nodes 3 and 9, they are  misclassified  most frequently. If the number of labeled data in each community
is increased to 2, the exact community structure can be got by both methods.
It is indicated that the more prior information ({\em i.e.} labeled vertices) we have,
the better the performance of SELP is.

\begin{table*}[!htp]
\centering \caption[Community detection results for the Karate Club network]{Community detection results for the Karate Club network.}~\\
\resizebox{\textwidth}{!}{
\begin{tabular}{rrrrrrrrrrr}
  \hline
 Labeled nodes in  $\omega_1$ & Labeled nodes in $\omega_2$ & Misclassified nodes by SELP & Detected outliers by SELP & Misclassified nodes by SLP \\
  \hline
1 & 34& None & 10, 12 & None\\
1 & 32& 9 & 10, 12 & 9, 10, 27, 31, 34\\
2 & 33& None & 10, 12 & None\\
6 & 31 &  3 & 10, 12 & 2, 3, 8, 14, 2 \\
8 & 31 & None & 10, 12 & 10\\
8 & 32  & None  & 10, 12 & None\\
17 & 31 & 3, 4, 8, 14 & 10, 12 & 2, 3, 4, 8, 13, 14, 18, 20, 22\\
1, 2 & 33, 34 & None & 10, 12 & None\\
1, 2 & 33, 9 & None & 10, 12 & None\\
3, 18 & 26, 30 & None & 10, 12 & None\\
17, 4  & 31, 9 & None & 10, 12 & None\\
   \hline
\end{tabular}}\label{karate_table}
\end{table*}

\textbf{B. American football network.}
As a further test of our algorithm,  the network we investigated in this experiment was the world of American college football games. %between
%Division IA colleges during regular season Fall 2000. The vertices  in the network represent 115 teams,
%while the links denote 613 regular-season games between the two teams they
%connect. The teams are divided into 12 conferences  containing around 8-12
%teams each and generally games are more frequent between members from the same
%conference than between those from different conferences.

\begin{center} \begin{figure}[!htp] \centering
		\includegraphics[width=0.45\linewidth]{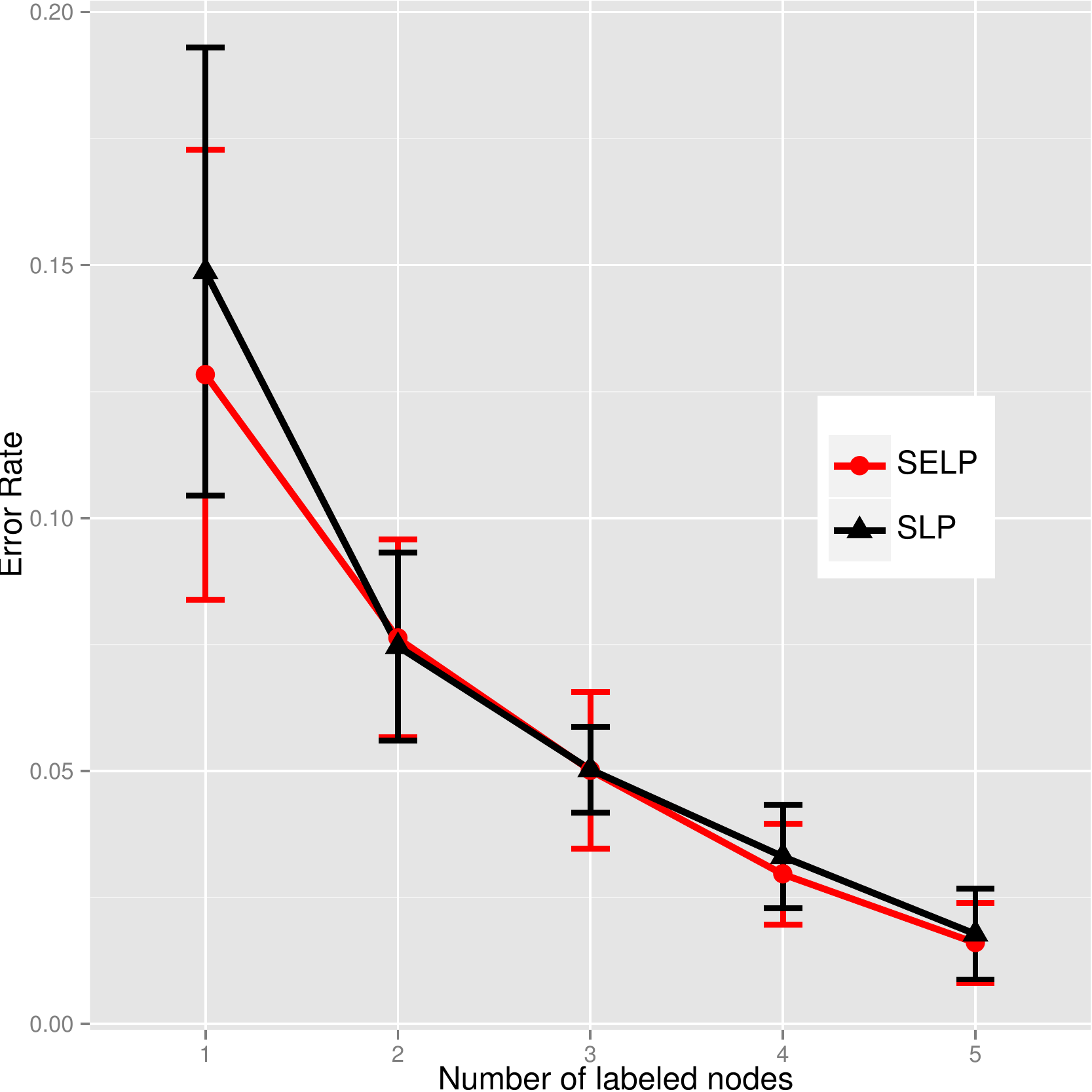}\hfill
        \includegraphics[width=0.45\linewidth]{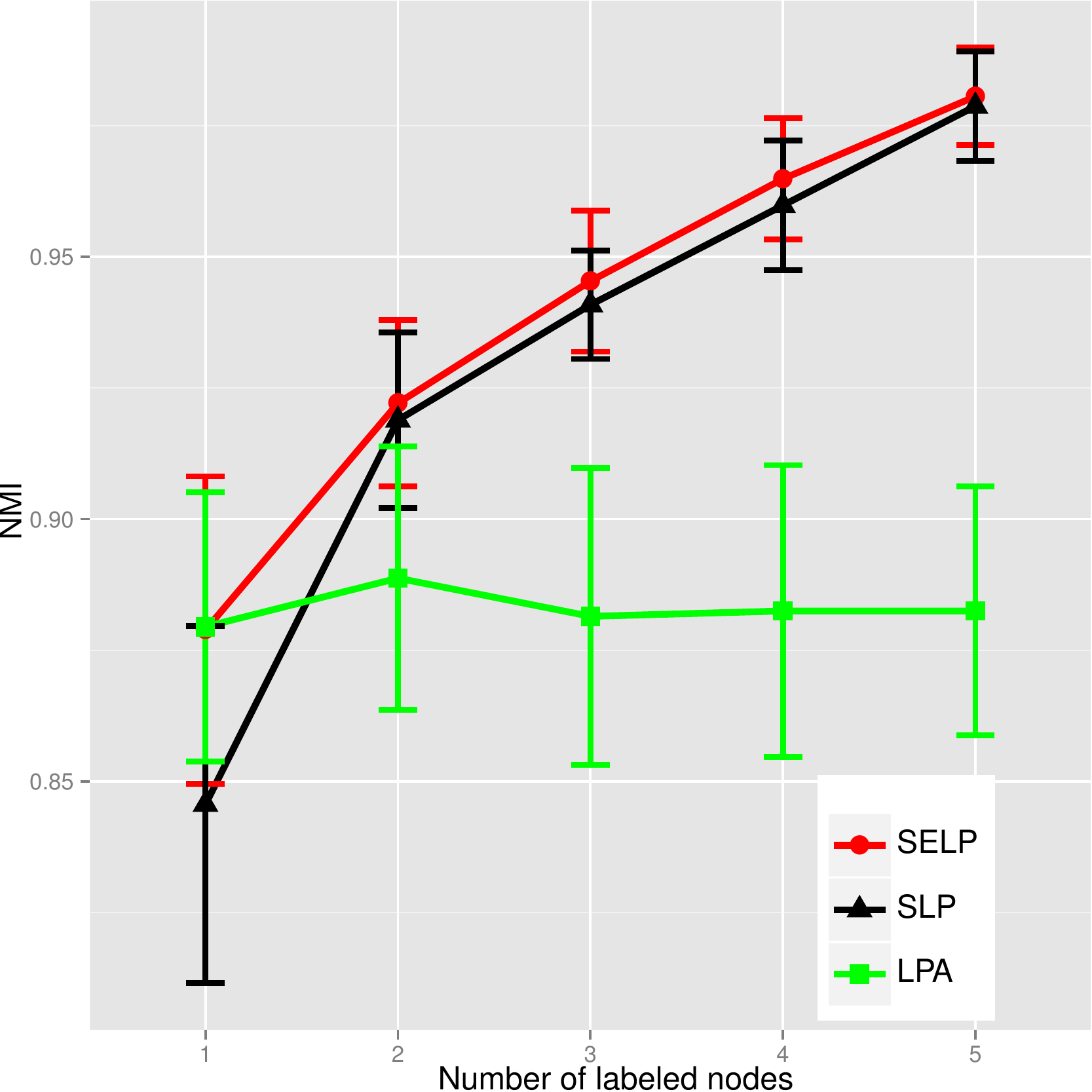} \hfill
        \parbox{.45\linewidth}{\centering\small a. Error rate} \hfill
		\parbox{.45\linewidth}{\centering\small b. NMI}
\caption[The results on American football network]{The results on American football network. The two
figures show the average error rates and NMI values (plus and minus one standard deviation) for 50 repeated
experiments, as a function of the  number of labeled samples.} \label{footballi_label} \end{figure} \end{center}

\vspace{-3em}
Let the number of labeled nodes in each community to be fixed. Then SELP and SLP algorithms were evoked 50 times respectively with randomly
selected labeled nodes. The average error rates and NMI values (plus and minus one standard deviation)
of the 50 experiments are displayed in Figures \ref{footballi_label}-a and \ref{footballi_label}-b respectively. As can be seen from the figures,
with  the increasing number of labeled samples, the performance of both SELP and SLP becomes better. The NMI values of the detected communities
by SELP and SLP are significantly better than those by LPA. It indicates that the semi-supervised community detection methods could take
advantage of the limited amount of prior information and consequently improve the accuracy of the
detection results. The behavior of SELP is better than that of
SLP in terms of both error rates and NMI values.

\subsection{LFR network}
In this subsection, LFR benchmark networks were used to test the ability of the algorithm to identify communities.
The experiments here included evaluating the performance
of the algorithm with various amounts of labeled nodes and  different values of parameter $\mu$ in the benchmark
networks. The original LPA \cite{raghavan2007near} and the semi-supervised community detection approach SLP \cite{liu2014semi} were used to compare.

In LFR networks, the mixing parameter $\mu$ represents the ratio between the
external degree of each vertex with respect to its community and the total degree of
the node. The larger the value of $\mu$ is, the more difficult the community structure
will be correctly detected.  The values of the parameters
in  LFR benchmark networks in this experiment were set as follows: $n = 1000, \xi = 15, \tau_1 = 2, \tau_2 = 1, cmin = 20, cmax = 50$.
\begin{center} \begin{figure}[!htp] \centering
		\includegraphics[width=0.45\linewidth]{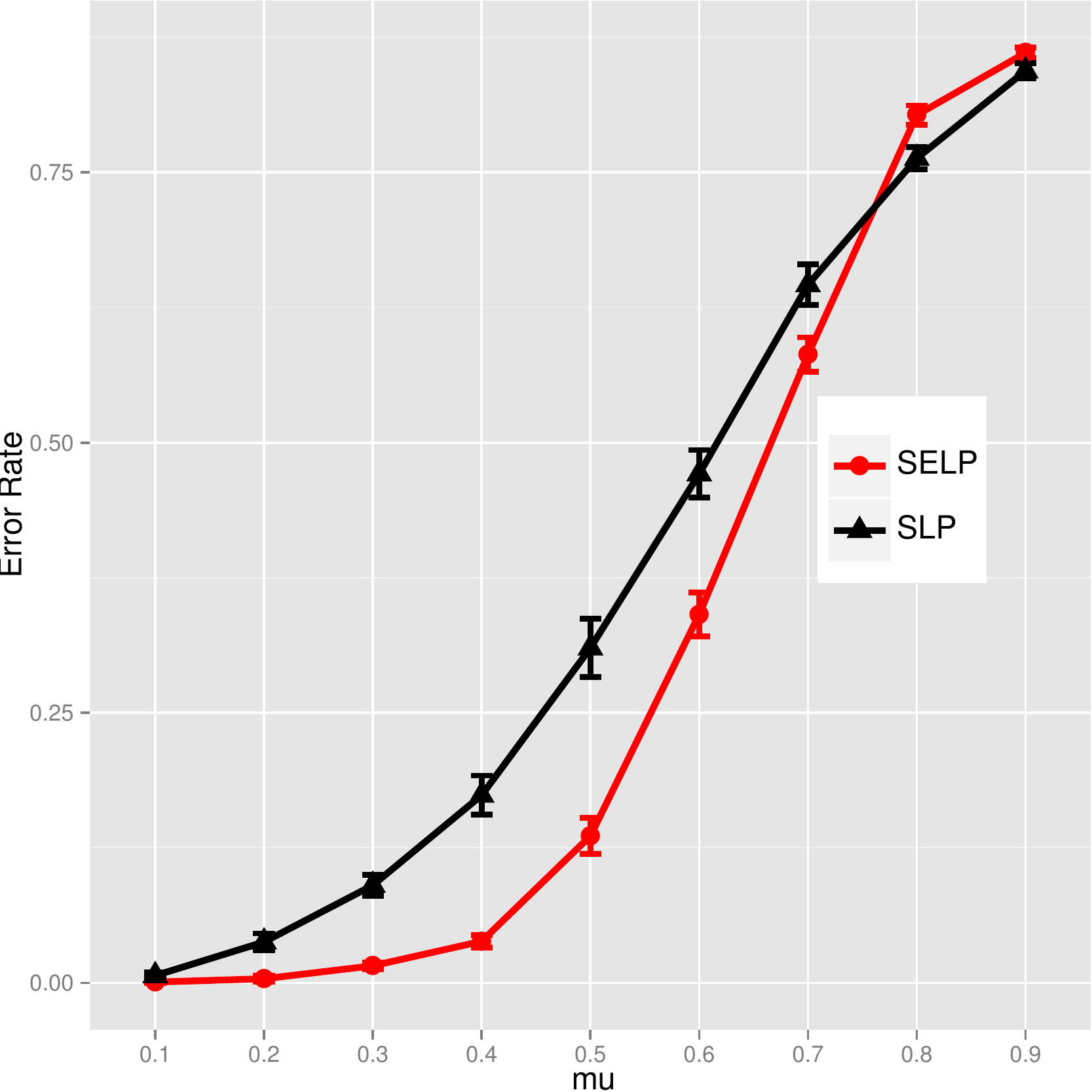}\hfill
        \includegraphics[width=0.45\linewidth]{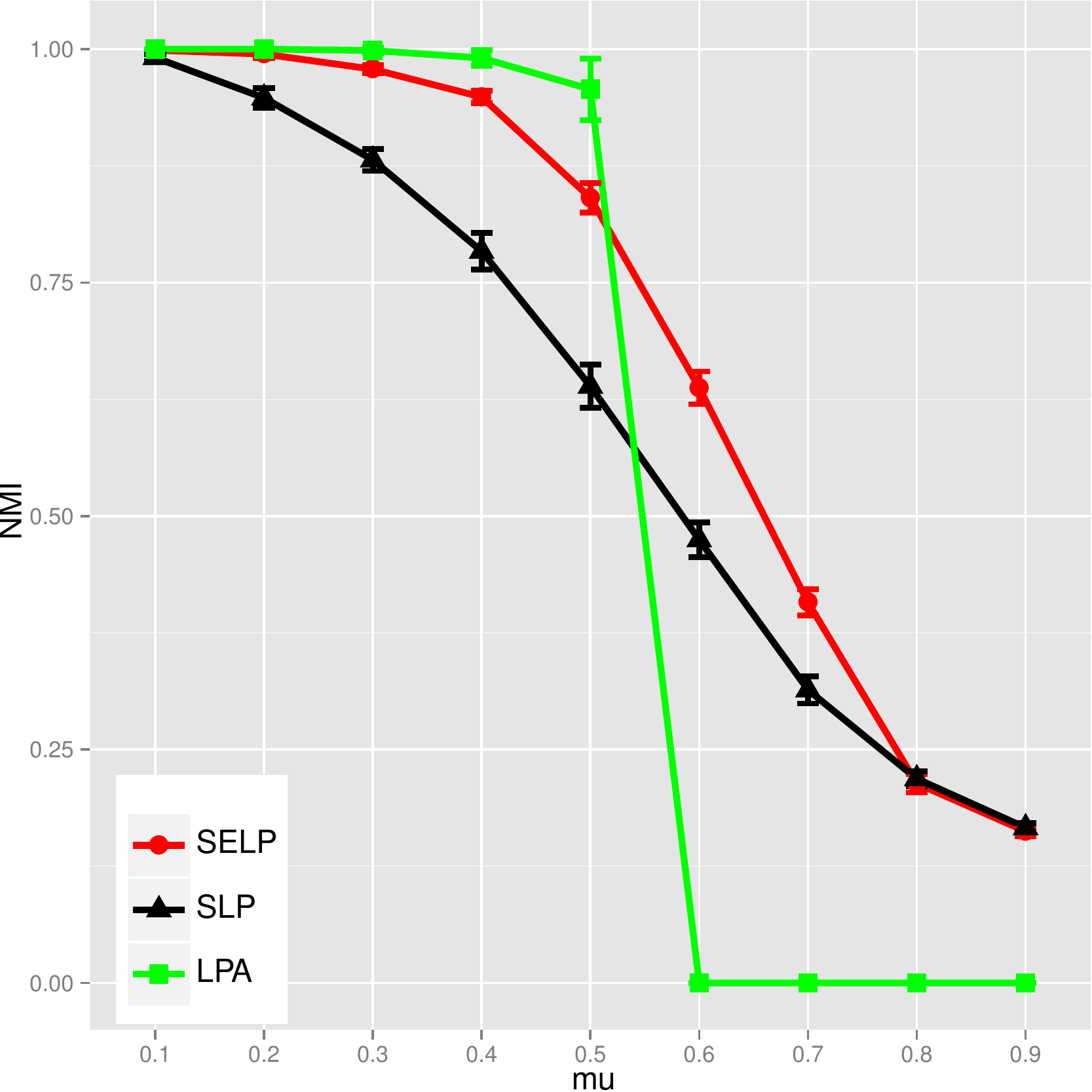} \hfill
        \parbox{.45\linewidth}{\centering\small a. Error rate} \hfill
		\parbox{.45\linewidth}{\centering\small b. NMI}
\caption[The results on LFR network]{The results on LFR network. The number of labeled nodes in each community is 3.} \label{LFR_mu} \end{figure} \end{center}

\vspace{-3em}
The performance of different methods with various values of $\mu$ is shown in Figure \ref{LFR_mu}.  As expected,
the error rate is very high and the NMI value is low when $\mu$ is large. It demonstrates the fact that
the community structure is not very clear and consequently difficult to be identified correctly. It can be seen
from Figure \ref{LFR_mu}-a that the error rates  by SELP are smaller than those by SLP generally.
SELP performs better than SLP. This conclusion could also be got in terms of
the NMI values displayed in Figure \ref{LFR_mu}-b.

\begin{center} \begin{figure}[!htp] \centering
		\includegraphics[width=0.45\linewidth]{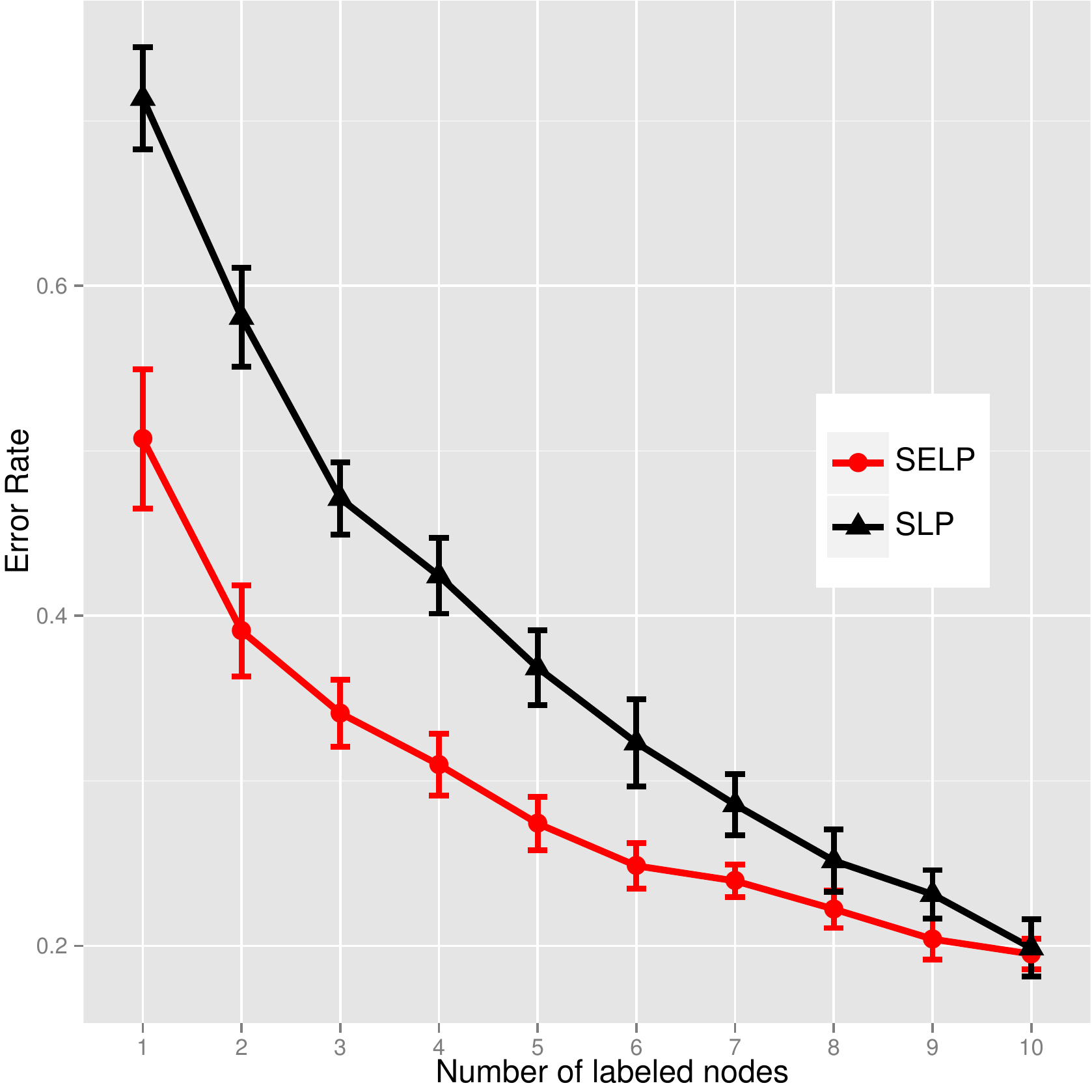}\hfill
        \includegraphics[width=0.45\linewidth]{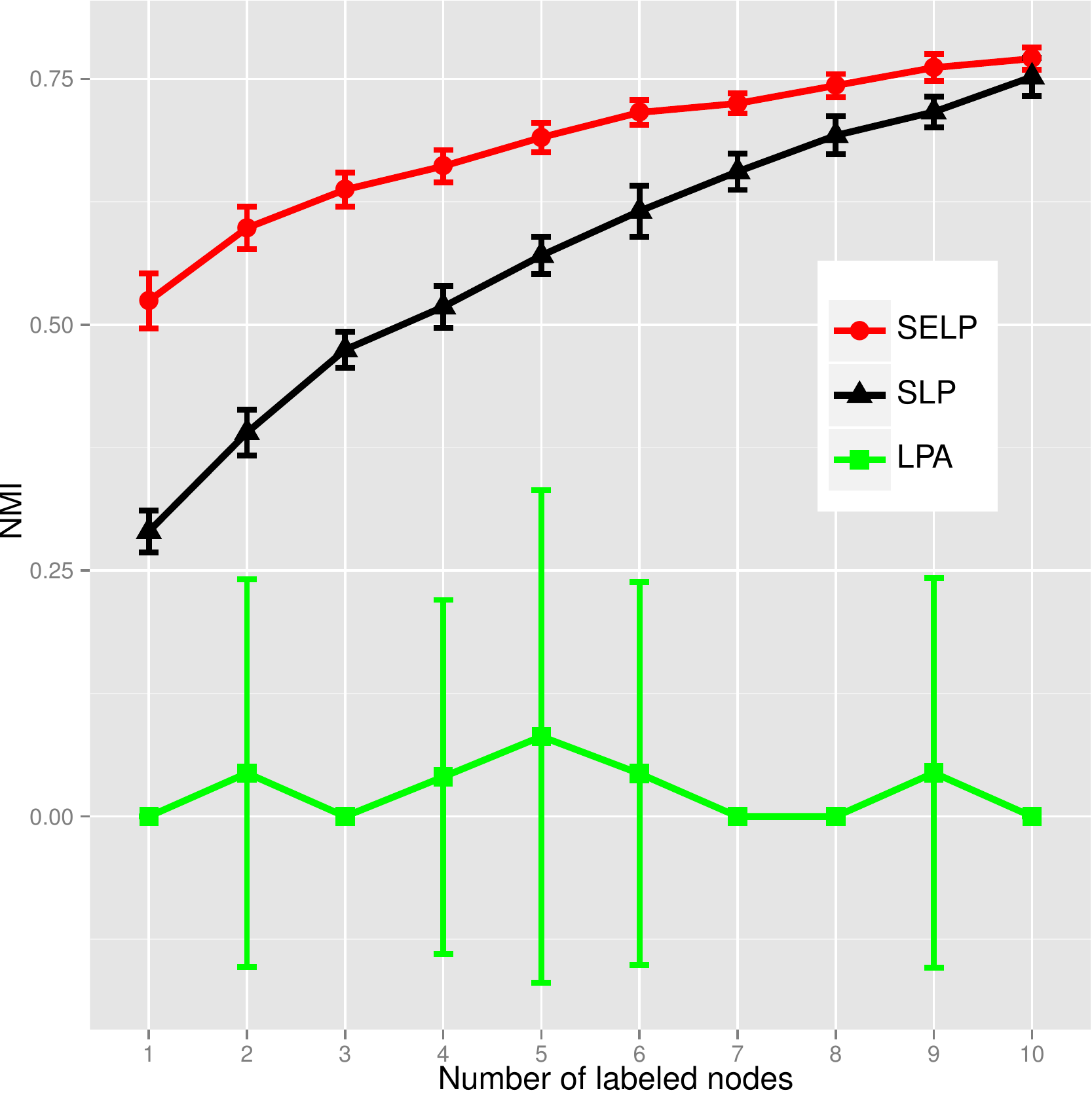} \hfill
        \parbox{.45\linewidth}{\centering\small a. Error rate} \hfill
		\parbox{.45\linewidth}{\centering\small b. NMI}
\caption[The results on LFR network]{The results on LFR network. The parameter of $\mu$ is set to be 0.6.} \label{LFR_labelX} \end{figure} \end{center}

The original LPA could not work at all when $\mu$ is larger than 0.5. The
results of SELP and SLP are significantly improved in these cases compared with LPA. As shown in Figure \ref{LFR_labelX}-b, even when there is
only one labeled data in each community, the behavior of SELP is much better than that of LPA. This  confirms the fact that
the semi-supervised community
detection approaches can effectively take advantage of the limited amount of labeled data.  From Figure \ref{LFR_labelX}, we can also see that
the performance of  SELP and SLP becomes better with the increasing number of labeled nodes.

\vspace{-5.3em}
\section{Conclusion}
In this paper, the semi-supervised evidential label propagation algorithm is presented as an enhanced version
of the original LPA. The proposed community detection approach  can effectively
take advantage of the limited amount of supervised  information. This is of practical meaning in real applications as
there often exists some prior knowledge for the analyzed graphs. The experimental results show that
the detection results will be significantly improved with the help of  limited amount of supervised information.

\bibliographystyle{splncs03}
\bibliography{paperlist}
\end{document}